\documentstyle[12pt]{article}

\def\ap{A(\vec p_1+\vec p_2)}
\def\am{A(\vec p_1-\vec p_2)}
\def\bp{B(\vec p_1+\vec p_2)}
\def\bm{B(\vec p_1-\vec p_2)}
\def\bd#1#2{#1\!\cdot\!#2}
\def\Pi2{\frac{\pi}2}

\def\be{\begin{equation}}
\def\ee{\end{equation}}
\def\bea{\begin{eqnarray}}
\def\eea{\end{eqnarray}}
\def\ba{\begin{array}}
\def\ea{\end{array}}
\def\bma{\left(\begin{array}}
\def\ema{\end{array}\right)}

\begin{document}

\title{Continuous vacua in bilinear soliton equations}

\author{
J. Hietarinta\thanks{permanent address:
Department of Physics, University of Turku,
20500 Turku, Finland}
\  and
A. Ramani\\
CPT, Ecole Polytechnique\\
91128 Palaiseau, France\\
and\\
B. Grammaticos\\
LPN, Universit\'e Paris VII, Tour 24-14, 5\`eme \'etage\\
2 Place Jussieu, 75251 Paris, France}

\date{October 26, 1993}

\maketitle

\begin{abstract}
We discuss the freedom in the background field (vacuum) on top of
which the solitons are built.  If the Hirota bilinear form of a
soliton equation is given by $A(D_{\vec x})\bd GF=0,\, B(D_{\vec
x})(\bd FF - \bd GG)=0$ where both $A$ and $B$ are even polynomials in
their variables, then there can be a continuum of vacua, parametrized
by a vacuum angle $\phi$.  The ramifications of this freedom on the
construction of one- and two-soliton solutions are discussed.  We
find, e.g., that once the angle $\phi$ is fixed and we choose
$u=\arctan G/F$ as the physical quantity, then there are four
different solitons (or kinks) connecting the vacuum angles $\pm\phi$,
$\pm\phi\pm\Pi2$ (defined modulo $\pi$). The most interesting result
is the existence of a ``ghost'' soliton; it goes over to the vacuum in
isolation, but interacts with ``normal'' solitons by giving them a
finite phase shift.
\end{abstract}

\section{Introduction}
The existence of multisoliton solutions has been considered as a very
strong indication of complete integrability of nonlinear evolution
equations.  The main tool for this was created by Hirota, who has
proposed his bilinear formalism based on the observation that,
expressed in the right variables, soliton solutions are just
polynomials in exponentials [1].  Indeed, when one starts from the
solitary wave solution
\be
u =\frac{k^2/2}{[\cosh((kx-pt)/2)]^2},\, k^3-p=0,
\ee
of the KdV equation
\be
u_t -6 u u_x + u_{xxx} = 0,\label{KdV}
\ee
and introduce the new dependent variable $F$ by $ u = 2 \partial _x^2
\log F $ one obtains simply $ F = 1 + e^{kx-pt}$.

The KdV equation itself can be cast in the bilinear form using the
new variable F. From the potential form of (\ref{KdV}), which is
$v_t - v_x{}^2+v_{xxx}=0$, where $v_x=u$, we find:
\be
F F_t + F F_{xxxx} - 4 F_x F_{xxx} + 3 F_{xx}{}^2 = 0,\label{BKdVa}
\ee
and using Hirota's $D$ operators defined through:
\be
D_x^a D_t^b \dots \bd FG =(\partial _x -\partial _x')^a
(\partial _t -\partial _t')^b \dots F(x,t,\dots) G(x',t',\dots) |_{x'=x,
t'=t, \dots},
\ee
we can write (\ref{BKdVa}) as
\be
(D_x^4 + D_x D_t ) \bd FF = 0.\label{BKdV}
\ee

Expressed in this new dependent variable $F$ the multisoliton
solutions of the KdV equation can be systematically
constructed (with $\eta =kx-pt$):
\begin{itemize}
\item[0):] zero soliton (vacuum): $ F = 1 $
\item[1):] one soliton : $ F = 1 + e^{\eta} $
\item[2):] two solitons : $ F = 1 + e^{\eta _1} + e^{\eta _2} +
A_{12} e^{\eta _1+\eta _2} $ with $ A_{12}=\frac{k_1-k_2}
{k_1+k_2} $
\item[3):] three solitons : $ F = 1 + e^{\eta _1} + e^{\eta _2}
+ e^{\eta _3} + A_{12} e^{\eta _1+\eta _2}
+ A_{13} e^{\eta _1+\eta _3} + A_{23} e^{\eta _2+\eta _3}
+ A_{12} A_{13} A_{23} e^{\eta _1+\eta _2+\eta _3} $
\end{itemize}
\noindent
etc.

We remark that the three-solitons solution does not contain
any new free parameters, i.e. once $A_{ij}$ is fixed at the two-%
soliton level the parameters of the three- (and higher) soliton
solutions are fixed. Thus, for the existence of these higher order
solutions it is necessary that certain  compatibility conditions be
satisfied.

Returning to (\ref{BKdV}) we emphasize that there exists a solution
corresponding to a `vacuum', i.e.\ the absence of solitons. One can
see that $F=$ constant $\ne 0$ is a solution which (due to
bilinearity) can be normalized to $F=1$. Thus, the vacuum in the case
of KdV is uniquely defined.

However, the vacuum solution is not always unique.  In the following
we will show that for certain multicomponent bilinear equations there
is a continuum of vacua which cannot be scaled away.  In this case the
solitons turn out to interpolate between different vacua, and there is
one instance where the soliton is a ``ghost'': It is not visible
alone, but only in interaction with others.  This multiplicity of
vacua puts also extra constraints on the existence of soliton
solutions and thus on the integrability of the equation.

\section{Continuous vacua for two-component bilinear equations}
Bilinear equations can be classified according to the number of
dependent functions one has to introduce in order to write them under
Hirota's form. Following the analysis in [2], we concentrate on two
generic classes of two-component bilinear equations which write:
\be
\left\{\begin{array}{lcr}
A(D_{\vec x}) \bd GF &=& 0,\nonumber \\
B(D_{\vec x}) \bd GF &=& 0,
\end{array}\right.
\ee
where $A$ and $B$ are respectively even and odd polynomials in the $D$
operator (the modified-KdV family) and
\be
\left\{\begin{array}{rcl}
A(D_{\vec x}) (\bd FF - \bd GG) &=& 0, \\
B(D_{\vec x}) \bd FG &=& 0,
\end{array}\right. \label{EEE}
\ee
where $A$ and $B$ are both even (the sine-Gordon family).

The even-odd case is `standard' in that only usual vacua exist.
We can write $F=s,\, G=c$, but due to bilinearity we can scale this
to $F=1,\,G=1$, or $F=1,\,G=0$, or $F=0,\,G=1$ so that no continuous
parameters remain.

Let us now focus on the even-even case (\ref{EEE}), and look for the
zero-soliton solution.  Two types of standard vacua can readily be
found.  If $A(0)=0$, $B(0)\ne0$ then $G=0$, $F=1$ and $F=0$, $G=1$ are
the only possible vacua.  Also if $A(0)\ne0$, $B(0)=0$ we must take
$F=G=1$ or $F=-G=1$.  Note that if neither $A(0)$ nor $B(0)$ vanish
there are no simple zero-soliton solution at all.

A novel possibility exists if both $A(0)$ and $B(0)$ vanish in
(\ref{EEE}).  Then if $F$ and $G$ are both constants, one gets a
vacuum solution, whatever their ratio.  We can use a symmetric
parametrisation of the continuous vacuum solution through
\be
F=\cos\phi,\quad G=\sin\phi,\label{Evac}
\ee
with $\phi$ a free parameter.

The construction of a one-soliton solution on top of one of the standard
vacua is straightforward and has been discussed e.g.\ in [2].
The question we will now address is whether one can find one-soliton
solutions on top of the new vacuum (\ref{Evac}). Let us start with the
ansatz:
\be
F = c + C e^\eta, \quad G = s + S e^\eta, \label{Ean}
\ee
where $c=\cos \phi ,\,s=\sin \phi,$ and $\eta =kx-pt+$const.
Substituting into (\ref{EEE}) we obtain the following conditions:
\be
B(\vec p) (sC+cS)=0,\quad A(\vec p) (cC-sS) =0,
\ee
where $\vec p=(k,p)$.  Three cases can be distinguished, corresponding
to three different types of solitons:
\begin{itemize}
\item[$\alpha)$] $A(\vec p)=0$ and $B(\vec p)\ne 0$ then $C=sM,\, S=cM$
 for some constant $M$.  This constant can be absorbed into the constant
in $\eta$, so that
\be
F = c(1+e^\eta ),\quad G=s(1-e^\eta ). \label{Eal}
\ee
\item[$\beta$)] $A(\vec p)\ne 0$ and $B(\vec p)=0$ then
\be
F = c+s e^\eta , G=s+c e^\eta.\label{Ebe}
\ee
\item[$\gamma$)] $A(\vec p)=0$ and
$B(\vec p)=0$ then there is, at this level, no condition on $C$ and $S$
and the one-soliton solution seems to be arbitrary.
\end{itemize}
\noindent
A typical situation where a $\gamma$-type soliton exists is when the
Hirota polynomials $A$ and $B$ have a common factor $U$: $ A=UV ,\,
B=UW $ and thus the dispersion manifold common to both $A$ and $B$
contains $ U(\vec p)=0$.  Moreover, in many cases, (in particular for
known integrable even-even systems), $U$ is just linear i.e.\  $
U(\vec p)= \bd{\vec \lambda}{\vec p} $.  This will have further
implications on the 2-solitons solution as we will see in the next
section.

\section{Constraints from the existence of two-soliton solutions}
In addition to the continuous parametrization of the vacuum there is a
further discrete ambiguity.  We can observe this already from
(\ref{Eal},\ref{Ebe}) when we look at the physical quantity $u=\arctan
G/F$: It has different (constant) values at $\eta\to\pm\infty$, which
suggest different vacua there.  This will be discussed further in
Sec.\ 5, but before doing that let us see what the formal requirement
for the existence of two-soliton solution yields.

Let us start by combining solutions of the types $\alpha$ and $\beta$
above.  First, for $\alpha +\alpha$ (which means that the dispersion
relations are $ A(\vec p_1)=A(\vec p_2)=0 $ for both of the individual
solitons), the ansatz for $F$ and $G$ write:
\bea
F&=& c+ce^{\eta _1}+ce^{\eta _2}+Ke^{\eta _1+\eta _2},\nonumber \\
G&=& s-se^{\eta _1}-se^{\eta _2}+Le^{\eta _1+\eta _2}.
\eea
and then we obtain from (\ref{EEE}) $K=cM,\,L=sM$ with
\be
M = -\frac{\am}{\ap}=\frac{\bm}{\bp}. \label{Eform}
\ee
This implies in particular a condition for the existence of the
two-solitons solutions: On the manifold $A(\vec p_1)=A(\vec p_2)=0$ we
must have:
\be
 \ap \bm + \am \bp =0. \label{Econd}
\ee
This condition reflects the fact that generically even-even
bilinear equations do not have two-solitons solutions.

The $\beta +\beta$ solution leads to analogous results:
\bea
F&=& c+s e^{\eta _1}+s e^{\eta _2}-cMe^{\eta _1+\eta _2},\nonumber \\
G&=& s+c e^{\eta _1}+c e^{\eta _2}-sMe^{\eta _1+\eta _2}.
\eea
with $M$ given by (\ref{Eform}) and (\ref{Econd}) as the existence
condition, but now it should be understood on a different dispersion
manifold, namely $ B(\vec p_1)=B(\vec p_2)=0. $

The case $\alpha +\beta$ is treated in a similar way. We start from
\bea
F&=& c+c e^{\eta _1}+s e^{\eta _2}+Ke^{\eta _1+\eta _2},\nonumber \\
G&=& s-s e^{\eta _1}+c e^{\eta _2}+Le^{\eta _1+\eta _2}.\label{Ealbe}
\eea
on $ A(\vec p_1)=B(\vec p_2)=0 $ (or $ B(\vec p_1)=A(\vec p_2)=0 $ )
and find $K=-sM,\,L=cM$ with (\ref{Eform}) and (\ref{Econd}) as the
compatibility condition (on a still different dispersion manifold).
Thus for the standard type solitons $\alpha$ and $\beta$ the
construction of two soliton solutions is straightforward and leads to
the condition (\ref{Econd}) on a suitable manifold.

More interesting are the cases where a $\gamma$-type soliton is
involved. Let us combine one $\alpha$-type soliton with a $\gamma$-%
type one. The dispersion manifold in this case is $ (A(\vec p_1)=0)
\cap (A(\vec p_2)=0) \cap (B(\vec p_2)=0)$ or $ (A(\vec p_1)=0)
\cap (B(\vec p_1)=0) \cap (A(\vec p_2)=0)$. Let us choose
the first case and write the solution as :
\bea
F&=&c+c e^{\eta _1} + C e^{\eta _2} + K e^{\eta _1+\eta _2},\nonumber \\
G&=&s-s e^{\eta _1} + S e^{\eta _2} + L e^{\eta _1+\eta _2}
\eea
We readily find that $K=CM$, $L=-SM$, with three possibilities for
$C$, $S$ and $M$, namely:
\begin{enumerate}
\item[1)] type $\gamma_1$: $C=c$, $S=s$, and $$ M=-\frac{\am}{\ap}. $$
\item[2)] type $\gamma_2$: $C=s$, $S=-c$, and $$ M=\frac{\bm}{\bp}. $$
\item[3)] type $\gamma_3$: $C$ and $S$ are free and $M$ given by
(\ref{Eform}) and therefore (\ref{Econd}) must be satisfied again on
the appropriate dispersion manifold.
\end{enumerate}
Similar conclusions can be reached in the $\beta +\gamma$ case,
mutatis mutandis.

Let us finally consider the interaction of two $\gamma$-type solitons.
Here, the dispersion manifold is : $(A(\vec p_1)=0) \cap (B(\vec p_1)=0)
\cap (A(\vec p_2)=0) \cap (B(\vec p_2)=0)$.  The two-soliton solution is
written
\bea
F&=&c+C_1 e^{\eta _1} +C_2 e^{\eta _2} +K e^{\eta _1+\eta _2},\nonumber \\
G&=&s+S_1 e^{\eta _1} +S_2 e^{\eta _2} +L e^{\eta _1+\eta _2}.\label{Egaga}
\eea
For notational convenience, we introduce the quantities $\varphi$
and $\psi$ through:
\be
K = s \varphi + c \psi,\quad L = c \varphi - s \psi,
\ee
and readily find
\bea
\varphi &=& -(C_1 S_2 + C_2 S_1) \frac{\bm}{\bp} ,\nonumber \\
\psi &=& -(C_1 C_2 - S_1 S_2) \frac{\am}{\ap}
\eea
provide that $\ap\ne0$ and $\bp\ne0$.

However, if the two $\gamma$-type solitons are obtained for $A=UV, \,
B=UW$ through $U(\vec p_1)=U(\vec p_2)=0$, and if moreover $U$ is
linear, then $U(\vec p_1\pm \vec p_2)=0$ as well and $A(\vec p_1\pm
\vec p_2)= B(\vec p_1\pm \vec p_2)=0$.  In that case with $F$ and
$G$ given by (\ref{Egaga}) one has $A(\bd FF-\bd GG)=B\bd FG=0$ for
arbitrary $C_i,\,S_i,\,K$ and $L$, and $\varphi$ and $\psi$ are still
totally free at this stage.

As we have seen above, the quantities $C_i$ and $S_i$ ($i=1,2$) are in
general fixed by the interaction of the $\gamma$-type solitons with
$\alpha$- and $\beta$-type ones, except in the special case of
$\gamma_3$ where (\ref{Econd}) would also be satisfied on the
appropriate dispersion manifold.  Thus if $\ap\ne0$ and $\bp\ne0$ the
only possible values of $C_i,\, S_i$ and consequently $\varphi$ and
$\psi$ are :
\begin{itemize}
\item[a)] $\gamma_1+\gamma_1$: $ C_1=C_2=c, S_1=S_2=s$ and
$$ \varphi = -2sc \frac{\bm}{\bp},\, \psi=(s^2-c^2)\frac{\am}{\ap}$$
\item[b)] $\gamma_1+\gamma_2$: $ C_1=c,C_2=s, S_1=s,S_2=-c$ and
$$ \varphi = (c^2-s^2)\frac{\bm}{\bp},\, \psi=-2sc\frac{\am}{\ap}$$
\item[c)] $\gamma_2+\gamma_2$: $ C_1=C_2=s, S_1=S_2=-c$ and
$$ \varphi = 2sc \frac{\bm}{\bp},\, \psi=(c^2-s^2)\frac{\am}{\ap}$$
\end{itemize}
In the exceptional case 3) above where $C$ and $S$ are free, the
$\gamma$-type soliton would remain completely free even at this
stage.

As the three- or more-soliton solutions exist only for integrable
systems, we cannot continue the analysis for general $A$ and $B$
polynomials.  The only known integrable cases for which $\gamma$-type
solitons exist at all do factorize with a linear $U$, so only the study
of the three-solitons solutions could determine the quantities $\varphi$
and $\psi$ which remain free at the two-soliton level.  This staggered
determination of the N-soliton solutions from the study of the (N+1)-one
was first remarked in our work [3] on ``static" solitons, i.e.  precisely
the case where the common factor $U$ is just $D_t$.

\section{Is there a free $\gamma$-type soliton for integrable systems?}
The fact that the parameters of a $\gamma$-type soliton may remain
undetermined even at the level of two-soliton solutions raises
questions about the nature of such a solution.  That is, should we
call this solution a `soliton' or not. The point is that it is in
general not true that any value of $C$ and $S$ in (\ref{Ean}) define a
{\it soliton}. For instance, let us assume that the common factor of
$A$ and $B$ is just $U=D_t$. Then, since $U$ is a factor of both $A$
and $B$ it is easy to convince oneself that any time-independent $F$
and $G$ will satisfy the equations.

However, not any time-independent object can be called a
time-independent soliton: there is the additional requirement that
upon interaction with any moving object it should re-emerge unchanged,
maybe up to a shift in position.  The condition for the $\gamma$-type
object defined above to satisfy this criterion is that it reduces to
one of the cases $\gamma_1$ ($C=c$ and $S=s$) or $\gamma_2$ ($C=s$ and
$S=-c$) defined above, unless condition (\ref{Econd}) happens to be
satisfied on the appropriate dispersion manifold, in which case the
$\gamma $-type soliton is still free at this stage.  Now we should
note, that while (\ref{Econd}) is satisfied on the $\alpha +\alpha$,
$\alpha +\beta$ and $\beta +\beta$ manifolds for all the known
integrable equations, it is not satisfied on the $\alpha +\gamma$ or
$\beta +\gamma$ manifolds for those few known equations where
$\gamma$-type solitons exist.  This means that the $\gamma $-type
solitons are in fact fixed at this stage for the {\it integrable}
cases.

{}From the analysis of the conditions for the existence of the
two-solitons solutions we obtained the condition (\ref{Econd}) on
various dispersion manifolds depending on the soliton solutions under
consideration.  This condition is a very strong one.  To start with,
it constitutes a first necessary condition for integrability.  Thus,
whenever the partial differential equation under consideration has a
Hirota form (\ref{EEE}) that possesses a continuous vacuum (i.e.$
A(0)=B(0)=0$), condition (\ref{Econd}) above may serve as a first
check for the integrability of the equation.

To illustrate this, let us consider the even-even bilinear
equation with Hirota polynomials
\be
A=D_x^3 D_t + D_y D_t + a D_x^2,\quad B= D_x( D_t + b D_x).\label{EInt}
\ee
If we write condition (\ref{Econd}) on $ A(\vec p_1)=A(\vec p_2)=0$ we
find the necessary condition $a=b=0$.  The same condition is obtained
on the manifold $ A(\vec p_1)=0 $, $ k_2=0 $ (from the first factor of
$B$).  On the other hand, if one wants to satisfy $ B(\vec p_2)=0 $
through the second factor, i.e.\ $ p_2+ b k_2 = 0 $ then (\ref{Econd})
is never satisfied, even for $a=b=0$.  In this last special case, and
in that case only, however, (\ref{Econd}) is {\it not} needed for
integrability, as for $p_2=0,\,a=b=0$ both $A$ and $B$ vanish and we
have in fact a $\gamma_1$- or $\gamma_2$-type soliton.  Finally
$a=b=0$ is a necessary and sufficient condition for the existence of
two-soliton solutions of all possible types ($\alpha,\,
\beta,\, \gamma_1,\, \gamma_2$).  One has thus obtained quickly the
only integrable subcase of (\ref{EInt}) at the two-soliton level.  If
one instead uses only the standard-type vacuum, one has to go to three
soliton solutions (and in exceptional cases even to four soliton
solutions) in order to restrict the values of $a$ and $b$.

Conversely, we could use (\ref{Econd}) in order to derive the possible
forms of bilinear PDE's which would possess a two-solitons solution in
the presence of a continuous vacuum.  However, the general solution of
the functional equation (\ref{Econd}) under the constraints defining
the dispersion manifold, seems to be a formidable task, (in particular
because it may happen that for some special values of the parameters
the existence of a solution is not obtained through (\ref{Econd}), but
instead because a $\beta$-soliton reduces to $\gamma $-type, as
happened in the case of (\ref{EInt}) above).  Moreover, in [4] we have
used general arguments based on singularity analysis and derived all
the possible bilinear even-even PDE's that could be candidates for
integrable equations, and on the light of these results finding the
general solution of (\ref{Econd}) does not seem to be necessary.  In
conlusion, (\ref{EInt}) with $a=b=0$ is the only integrable pair with
a continuous vacuum (and $A\ne B$).

\section{The physical vacuum and solitons interpolating between them}
Let us now return to the question of the vacuum angle.  We noted
already that even when $\phi$ is fixed there are in fact several
`physical' vacua, and the soliton solutions connect them.  First
evidence of that is obtained when we recall that bilinear equations
are invariant under a simultaneous change of phase.  Thus (\ref{Ean})
can also be written as
\bea
 \tilde F:=e^{-\eta}F &=& e^{-\eta}c + C, \nonumber \\
\tilde G:=e^{-\eta}G &=& e^{-\eta}s + S,
\eea
so that the soliton seems to be build on top of the vacuum $\tilde
F=C,\,\tilde G=S$.

Further information on the vacuum is obtained when we use the physical
quantities. It is well known that $F$ or $G$ alone do not have
physical meaning as they blow up when $\eta\to\infty$. The typical
physical variable is
\be
u=\arctan(G/F),
\ee
so let us see how the vacua look like from the point of view of $u$.
The starting vacuum (\ref{Evac}) yields $u=\phi$, and that value is
obtained also from (\ref{Ean}) when $\eta\to-\infty$.  When
$\eta\to\infty$ we find that the limiting value is different for
different solitons:
\begin{itemize}
\item[$\alpha$:] $\phi\to-\phi$,
\item[$\beta$:] $\phi\to\Pi2-\phi$,
\item[$\gamma_1$:] $\phi\to\phi$,
\item[$\gamma_2$:] $\phi\to\phi-\Pi2$.
\end{itemize}
The solitons do therefore connect different values of the vacuum.
Note that from the point of view of $u$ the vacuum angle is defined
only modulo $\pi$.

When we look at the two-soliton solutions exhibited in Sec.\ 3 we
observe that it is really this change in the angle that characterizes
the soliton.  For example (\ref{Ealbe}) connects from vacuum angle
$\phi$ to $\phi-\Pi2$ and the intermediate vacuum angle is $-\phi$ or
$\Pi2-\phi$, depending on which order the $\alpha$ and $\beta$ kinks
appear (see Figure 1).  That is, the values of the soliton's leftside
and rightside vacuum angles may change during the interaction, but the
operation made on the vacuum angle will stay invariant and may be
associated with the soliton.  Figure 2 shows how the $\alpha$-soliton
interpolates between different vacua.

The soliton $\gamma_1$ is quite curious, because in isolation it goes
over to the vacuum.  It is therefore a kind of ``ghost'' soliton, and is
invisible when taken isolated.  However, when it interacts with a
normal soliton it manifests itself in an unambiguous way.  Figure 3
shows the time evolution of an $\alpha+\gamma_1$ pair and the effect
of the $\gamma_1$-soliton is only in a phase shift in the evolution of
the $\alpha$ soliton.

\section{Conclusions}
To conclude, we remark that the (continuous) vacuum multiplicity is an
interesting property of even-even bilinear equations which allows
surprising phenomena to occur. One of most remarcable is the existence
of hidden solitons which appear only in interaction with ``normal''
solitons.  For these systems it turns out that already the existence
of general two-soliton solutions can be used for investigating the
integrability in a simple manner.  Unfortunately, integrable systems
having this property are quite rare and few examples are known to
date.

\section*{Acknowledgements}
We would like to thank J.  Satsuma for comments on an early version of
the manuscript.

\section*{Figure captions}

\subsection*{Figure 1:}
The time evolution of the $\alpha+\beta$ soliton solution of (\ref{EInt})
with $a=b=0$ (the integrable case). The four different vacuum levels are
clearly visible. (For this equation the $\beta$-soliton is $x$-independent.)

\subsection*{Figure 2:}
This figure shows how the $\alpha$-soliton (\ref{Eal}) interpolates
different vacua. For figure a) we have assumed that $e^\eta$ in (\ref{Eal})
is positive, for figure b) that it is negative.

\subsection*{Figure 3:}
Contour lines for the time evolution of the $\alpha+\gamma_1$ soliton.
For $t\to\pm\infty$ only the $\alpha$ soliton is visible, the
$\gamma_1$ manifests itself only at the point where the $\alpha$
soliton goes over it.

\end{document}